# SMART BUT NOT MORAL? MORAL ALIGNMENT IN HUMAN–AI DECISION-MAKING

*TREO Paper*

Christiane Ernst, University of Innsbruck, Innsbruck, Austria, christiane.ernst@uibk.ac.at

Luis Gutmann, University of Innsbruck, Innsbruck, Austria, luis.gutmann@uibk.ac.at

Domenique Zipperling, University of Bayreuth and Fraunhofer FIT, Bayreuth, Germany, domenique.zipperling@uni-bayreuth.de

Kathrin Figl, University of Innsbruck, Innsbruck, Austria, kathrin.figl@uibk.ac.at

Niklas Kühl, University of Bayreuth and Fraunhofer FIT, Bayreuth, Germany, kuehl@uni-bayreuth.de

## 1   Introduction

Artificial intelligence (AI) is increasingly embedded in high-stakes human–AI decision-making contexts such as hiring (e.g., Dargnies et al., 2024). These decisions are not merely technical. They often lack a clear ground truth and instead involve fundamentally moral judgments about fairness, responsibility, and harm. As a result, AI systems are no longer perceived as neutral tools, but as actors that implicitly or explicitly embody moral positions (e.g., Bonnefon et al., 2024).

At the same time, a growing body of research shows that alignment between AI systems and human stakeholders shapes key outcomes such as perceptions, trust, and reliance and third-party acceptance of AI-supported decisions (e.g., Bhat et al., 2024; Kirshner, 2024). Prior research has identified several determinants of reliance on AI advice, including psychological distance (Kirshner, 2024) and bias alignment (Zipperling et al., 2025). However, its primary focus has been on functional or behavioral alignment, while largely neglecting moral alignment, which may represent a more fundamental and potentially more consequential dimension of human–AI interaction, particularly given that AI has been conceptualized as capable of representing and operationalizing human moral values (Bonnefon et al., 2024). This gap is critical. In real-world deployments, stakeholders evaluate AI systems not only in terms of accuracy but also in terms of their perceived moral appropriateness. We define moral alignment as the perceived congruence between the moral values reflected in an AI system's decision logic and the moral intuitions of a given stakeholder. If an AI system reflects underlying values that conflict with stakeholders' moral intuitions, even highly accurate recommendations may be less likely to be accepted. Conversely, systems perceived as morally aligned may still be trusted and followed despite imperfections.

Despite these advances, we still lack a clear understanding of how moral alignment between AI systems and human stakeholders shapes perception, trust, reliance, and decision acceptance. This is not a marginal issue, it is central to whether AI will be meaningfully integrated into sensitive domains such as human resource management (Dargnies et al., 2024). Therefore, this study addresses the following research question: *How does moral (mis)alignment between the AI system and human stakeholders shape perceptions, trust, reliance, and decision acceptance in human–AI decision-making?*

## 2     Understanding Moral Alignment Through Haidt's Moral Foundations

To systematically examine moral alignment, we draw on Moral Foundations Theory (MFT) as a guiding framework (Haidt & Graham, 2007). MFT provides a structured account of how individuals differ in their moral intuitions, making it particularly suitable for analyzing value-based alignment between humans and AI systems. A large body of research shows that value similarity is an important driver of trust and acceptance. Consistent with similarity-attraction theory, individuals tend to evaluate others, whether human or artificial, as more trustworthy when they perceive shared values (Byrne, 1972). More broadly, this pattern aligns with the logic of value homophily: perceived similarity fosters psychological closeness, affiliation, and, ultimately, a greater willingness to rely on others' judgments (McPherson et al., 2001). MFT posits that moral reasoning is grounded in a limited set of evolutionarily shaped psychological systems that differ across individuals and cultures (Graham et al., 2009): *Care* (harm avoidance), *Loyalty* (group affiliation), *Authority* (respect for hierarchy), *Purity* (avoidance of contamination), and *Fairness* (justice). Recent work has refined fairness into two dimensions: *Equality* (equal treatment/outcomes) and *Proportionality* (merit-based reward) (Atari et al., 2023).

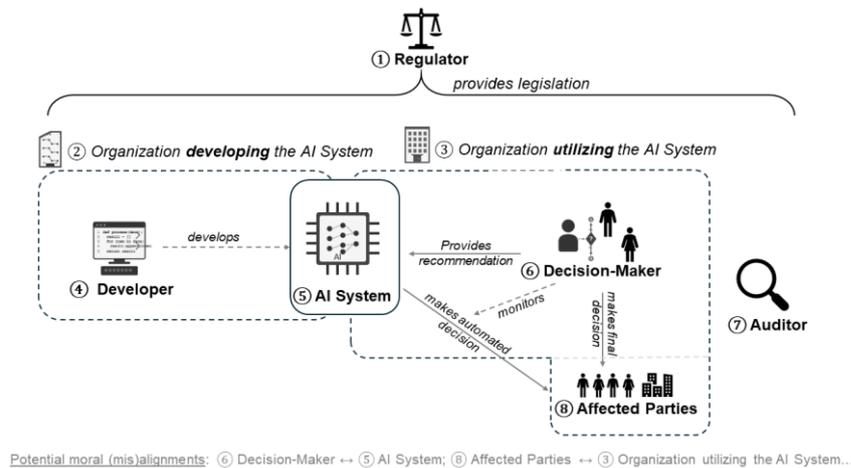

*Figure 1.     Stakeholders and potential moral (mis-)alignments in human–AI decision-making.*

Building on this framework, we conceptualize moral alignment not as a static property of an AI system, but as a relational construct. While regulatory frameworks such as the EU AI Act establish important normative boundaries—grounding AI systems in values such as human dignity and non-discrimination and requiring human oversight for high-risk systems—they do not account for how such systems are perceived by different stakeholders. Moral alignment emerges between specific stakeholders, such as decision-makers, organizations, and AI systems, whose underlying values may converge or conflict.

Figure 1 illustrates potential stakeholders and relationships of potential (mis-)alignment in human–AI decision-making contexts. Depending on the application domain, both the relevant stakeholders and the consequences of alignment or misalignment may vary substantially. This distinction is crucial: perceptions of moral alignment may vary not only across stakeholder groups, such as decision-makers, auditors, and affected parties, but also among individuals within the same group, as differences in moral priorities shape whether a given AI system is perceived as aligned or misaligned. Consequently, moral alignment is inherently context-dependent and shaped by the interplay of stakeholders involved in the decision-making process, resulting in a multi-stakeholder perspective.

## 3     Arising Challenges in Practice

This multi-stakeholder perspective implies that moral alignment cannot be assessed solely at the level of AI design. A practice-based illustration is recent public criticism of LinkedIn's content-ranking system, where users framed reduced reach as a moral concern about fairness, transparency, and access to opportunity rather than merely a technical issue (Evans, 2025). Though not definitive evidence of bias, such claims illustrate how perceived moral misalignment can turn AI-mediated systems into sites of stakeholder contestation. Even when developers seek to build systems that are morally aligned with the values or interests of affected parties and the organization, such moral alignment remains necessary but not sufficient. Its effects depend on how stakeholder relationships interact and how decision authority is distributed. Human oversight in AI-supported decision-making may therefore become a mechanism through which AI recommendations are discounted or overridden when they are perceived as morally misaligned with the moral intuitions of the human decision-maker. Consistent with prior work showing that individuals prefer AI systems aligned with their own predispositions (e.g., Zipperling et al., 2025), moral alignment in multi-stakeholder contexts cannot be equally realized for all stakeholders at the same time. Consequently, alignment with decision-makers may exert disproportionate influence on decision outcomes because their positional power enables them to interpret, adopt, or override AI recommendations.

This challenge points to three research avenues: **stakeholder constellations**, **types of moral conflict**, and **degrees of alignment**. Future research should therefore examine moral (mis-)alignment across developers, decision-makers, affected parties, auditors, and regulators, while varying both the moral values at stake and the extent to which AI systems align with different stakeholders. This shifts the focus from algorithm accuracy alone to the question of whose values are ultimately reflected in AI-supported decision-making in practice.